\documentclass{acm_sen_article}

\usepackage{lipsum}
\usepackage{flushend}
\usepackage{enumitem}
\usepackage{tabularx}
\usepackage{booktabs}
\usepackage{fontawesome5}
\usepackage{hyperref}
\usepackage{cleveref}

\usepackage[autostyle, english=american]{csquotes}
\MakeOuterQuote{"}

\usepackage{quoting}
\quotingsetup{leftmargin=0.5em, rightmargin=1em, vskip=0pt}

\begin{document}

\title{ICSE 2023 Sustainability Report}

\numberofauthors{5} 
\author{
Patricia Lago\\
       \affaddr{Vrije Universiteit Amsterdam}\\
       \affaddr{The Netherlands}\\
       \email{p.lago@vu.nl}
\and
Marcel Böhme\\
       \affaddr{Max Planck Institute (MPI-SP)}\\
       \affaddr{Germany}\\
       \email{marcel.boehme@mpi-sp.org}
\and 
Markus Funke\\
       \affaddr{Vrije Universiteit Amsterdam}\\
       \affaddr{The Netherlands}\\
       \email{m.t.funke@vu.nl}
}

\maketitle

\begin{abstract}
With growing discussions about the carbon footprint of academic conferences, more questions are being raised whether the environmental impacts caused by transportation and other factors justify the value of traditional paper presentations and social events. There is a pressing need to critically evaluate whether the ecological consequences of these events outweigh their perceived benefits. To that extent, we conducted a questionnaire survey among participants of the 45th International Conference on Software Engineering (ICSE) 2023 in Melbourne, Australia, seeking their feedback on the different conference sessions (e.g., workshops, keynotes, paper presentations, social events). In total, 161 participants filled out our survey. Overall, the conference was rated with 4.4 stars out of 5 stars. We do not see any significant differences among the different sessions, making it difficult to derive conclusions about their certain value and implications to sustainability. The relatively low response rate (11\%) did not help in gaining better insights.
Based on the participants registration data, we additionally estimated the carbon footprint emerged from air travel. The total carbon dioxide equivalent (CO2e) accumulates to around 5,053.5 tonnes of CO2e which is equivalent to the electricity required to power around 1,000 homes in a year.
With this report, we want to provide guidance to organizers of future conference editions with respect to their location and the perceived value of traditional paper presentations, social events, and other sessions. 
\end{abstract}


\section{Introduction}
In recent years, there has been a growing concern regarding the carbon footprint associated with academic conferences. The environmental impacts arising from transportation like long-haul flights or the energy consumed by the conference venue, have raised important questions about the justification of traditional paper presentations and social events in light of their ecological consequences. It is crucial to critically evaluate whether these events truly deliver significant benefits to participants that outweigh their environmental costs.

To better understand the perceived value of in-person attendance at large international conferences in light of their (environmental) impact, we started an exploratory investigation among attendees of our flagship software engineering conference---ICSE (International Conference on Software Engineering). For its 44th edition held in 2022 in Pittsburgh, USA, we conducted a post-conference questionnaire survey among participants seeking their feedback about the conference and experience from a sustainability perspective (see \cite{lago_icse_2026}). Our goal was to assess whether the benefits of in-person attendance would outweigh the carbon emissions generated. Overall, 8 of 42 respondents felt that the community's carbon footprint was not offset by the benefits of in-person attendance, while 15 respondents felt that the footprint was indeed offset. The remaining participants were unsure. By asking also open-ended questions, respondents mentioned mostly social factors as advantages of in-person meetings, such as networking, advertising their research, and providing feedback for early career researchers. However, due to the survey's low response rate, we continued our investigation for the next ICSE edition in 2023.

For the 45th edition of ICSE, held in Melbourne, Australia, we conducted another post-conference questionnaire survey among attendees aimed to collect feedback from participants regarding their perceptions of the various conference sessions and to assess the overall rating of the event. In addition, we also calculated the total carbon dioxide equivalent (CO2e) emerging from air traveling to and from the conference location in Melbourne. We estimate the CO2e based on the attendees’ registration data such as their indicated country.

The main objective of this report is to present the survey results and assess the perceived value of the ICSE 2023 conference, considering its huge environmental impact due to its location in Melbourne, Australia. Additionally, we explore potential differences in participant satisfaction among various session types. We want to provide guidance to organizers of future conference editions with respect to their location and the value of paper presentations and social events. Based on a first estimate of this year's emerging air travel related carbon footprint, we can provide historical data and a baseline that might be useful for comparison and further analysis for future editions of ICSE.

\section{Method}
\subsection{Questionnaire Survey}
To design our questionnaire we used Google Forms\footnote{Google Forms - \url{https://www.google.com/forms/about/}} as a survey tool. To keep the survey as concise as possible, our questionnaire encompassed only three questions as outlined in \Cref{tab:questionnaire}. To gain further insights between social events and traditional paper presentation sessions, we provided a list of 13 predefined session types, defined as:

\begin{itemize}[itemsep=0.1em, topsep=0em]
  \item Workshop
  \item Co-located event
  \item Keynote session (ICSE main)
  \item Paper-presentation session (ICSE main)
  \item EDI Welcome Reception
  \item Newcomers Reception
  \item Yoga Session
  \item Women@ICSE Lunch
  \item Conversation Café’ EDI
  \item Conversation Café’ Wellness in Academia
  \item Book club
  \item Lunch for LGBTQ+
  \item Other (add to your feedback the session type if not included before)
\end{itemize}
The feedback was collected by distributing printed QR codes (linking to the online questionnaire) at the venue and placing them near the doors of the meeting rooms.

\begin{table*}[h]
\centering
\caption{Questionnaire definition}
\label{tab:questionnaire}
\renewcommand{\arraystretch}{1.3}
\begin{tabularx}{\linewidth}{XXX}
\toprule
\textbf{Question} & \textbf{Type} & \textbf{Motivation} \\
\midrule
\textit{"Session Type"} & pre-defined list as specified & capture the session type to gain further insights between the different types \\

\textit{"How would you rate this session?"} & likert-scale from 1-to-5 stars (visualized through stars \faIcon{star}) & derive insights into the ranking of the different sessions \\

\textit{"Any other feedback?"} & open text field & gain further insights \\
\bottomrule
\end{tabularx}
\end{table*}

\begin{figure*}
    \centering
    \includegraphics[width=0.7\linewidth]{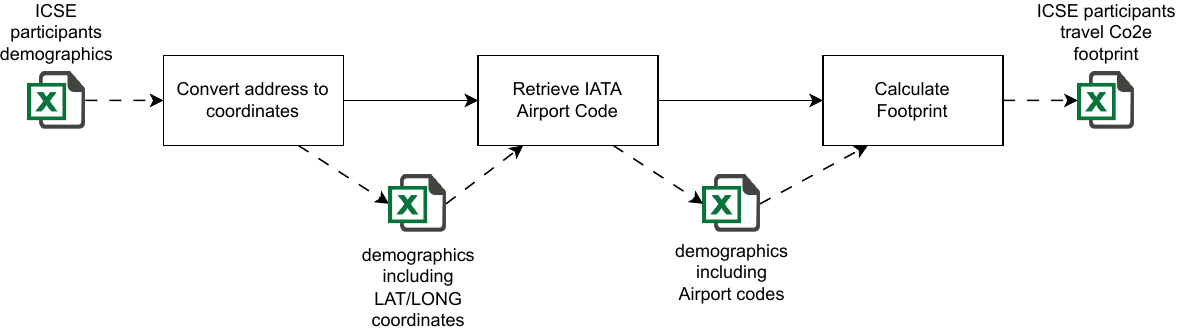}
    \caption{Carbon footprint calculator pipeline}
    \label{fig:pipeline}
\end{figure*}

\subsection{Carbon Footprint}
The registration period was already closed as it was decided to calculate the carbon footprint for the air travel related activity of our attendees. Therefore, we used the existing registration data to calculate the emerging footprint \textit{post-conference}. Due to the nature of the used registration form, the attendees origin was captured as basic free-text which allowed different spellings (e.g., "Chofu-shi, Tokyo-to, Japan" versus "Chofu-si, Tokyo, Japan"). Moreover, based on the registration data, it is also not evident whether an address does also represent the origin of the participants’ journey. Hence, we define our footprint calculation as an estimation as we took several assumptions. Various intermediate steps were necessary to calculate the CO2e for one single trip. We developed an internal carbon footprint calculator based on external APIs to overcome the aforementioned obstacles. The high level overview of the calculation pipeline is depicted in \Cref{fig:pipeline}.

\paragraph{1) Convert address to coordinates}
As mentioned, we first had to clean the derived registration demographics for similar spellings, typos, or unnecessary input. Basic Python operations were used for this exercise. Most online carbon footprint APIs require the request in the form of IATA airport codes\footnote{IATA Codes - \url{https://www.iata.org/en/publications/directories/code-search/}}. Therefore, a conversion from the available address into an airport code is necessary. However, to retrieve the nearest airport, the address needs to be translated into geographic coordinates, i.e., latitude and longitude. We use the Open Weather API\footnote{Open Weather API - \url{https://openweathermap.org/api}} to calculate the coordinates (lat/long) of attendees' anonymised addresses using city and country names. The accuracy here can vary and in some cases manual edit was required, as it was the case for some US states.

\paragraph{2) Retrieve IATA Airport Code}
We use the Airlabs API\footnote{Airlabs API - \url{https://airlabs.co/docs/airports}} to calculate the closest airport to the aforementioned coordinates within 100 km. This will convert the coordinates to IATA airport codes. Such a conversion may result in airports that were not actually selected by the participant, either because they were not available for international flights, or because they chose a different airport for other reasons. Therefore, the IATA airport code retrieved is only an approximation of the address entered by the respondent.

\paragraph{3) Calculate Footprint}
The Climatiq API\footnote{Climatiq - \url{https://www.climatiq.io/}} has been used to calculate the actual CO2e to Melbourne. Most international flights, especially those originating in Europe, include at least one stopover and can therefore be considered multi-leg flights. However, as our participants' stopovers cannot be recovered, our calculation is based on direct flights to Melbourne only. As most CO2e is generated during the take-off and landing phase, our assumption of single-leg flights could result in significantly lower emissions overall.

\section{Results}
This section presents an analysis of the survey results, focusing on participant ratings of the conference as a whole and their views on the different session types. Despite the overall success of the 45th edition of the ICSE conference with 1.426 attendees in total, our survey got a low response rate of 11.3\% with 161 participants in total. The final survey data and analysis are available online\footnote{ Survey data - \url{https://docs.google.com/spreadsheets/d/19OJAX8roRYJQ3IcbdwJThSosBev06P4IvKYX8YhWSQA}}.  According to the analysis in Table 2, overall, the participants rated the conference with 4.4 stars on average (arithmetic mean).

Unfortunately, the questionnaire did not distinguish between participants who attended on-site and remote. If we look at the open-ended question, there are some answers leading to an unclear picture. While some attendees experienced the remote setting as good:

\begin{quoting}
"I was one of the remote presenters @ AST. I think the org team did a fantastic job setting up the remote presentation option through zoom. The instruction we received in advance were very helpful." \\ \textbf{Session}: co-located event; \textbf{Stars}: 5
\end{quoting}

others considered the remote setting as sub-optimal:

\begin{quoting}
"Drop video presentations!! And there where no questions :-(" \\ \textbf{Session}: co-located event; \textbf{Stars}: 1
\end{quoting}

\begin{table*}[h]
\centering
\caption{Survey results grouped by the session type and their rating. The average represents the arithmetic mean.}
\label{tab:surveyresults}
\renewcommand{\arraystretch}{1.2}
\begin{tabularx}{51em}{>{\raggedright\arraybackslash}p{20em}
                              >{\centering\arraybackslash}p{1.2em}
                              >{\centering\arraybackslash}p{2.4em}
                              >{\centering\arraybackslash}p{3.6em}
                              >{\centering\arraybackslash}p{4.8em}
                              >{\centering\arraybackslash}p{6.0em} 
                              | >{\centering\arraybackslash}p{1.2em} 
                               >{\centering\arraybackslash}p{3.5em}}
\toprule
\textbf{Session type} & \faIcon{star} & \faIcon{star}\faIcon{star} & \faIcon{star}\faIcon{star}\faIcon{star} & \faIcon{star}\faIcon{star}\faIcon{star}\faIcon{star} & \faIcon{star}\faIcon{star}\faIcon{star}\faIcon{star}\faIcon{star} & \textbf{Total} & \textbf{Avg} \\
\midrule
co-located event                        & 2 & 1 & 3 & 15 & 16 & 37 & \textbf{4.14} \\
keynote                                 &   &   &   &   & 7  & 7  & \textbf{5} \\
keynote session (ICSE main)            & 1 & 2 & 1 &   & 2  & 16 & \textbf{4.13} \\
other                                   &   &   &   &   & 5  & 5  & \textbf{5} \\
paper-presentation                      &   &   & 1 & 3 & 13 & 17 & \textbf{4.71} \\
paper-presentation session (ICSE main) & 1 & 2 & 6 & 11 & 24 & 44 & \textbf{4.25} \\
women@ICSE lunch                        &   &   &   & 3 & 9  & 12 & \textbf{4.75} \\
workshop                                &   &   &   & 3 & 20 & 23 & \textbf{4.87} \\
\midrule
\textbf{Grand Total}                    & \textbf{4} & \textbf{5} & \textbf{11} & \textbf{37} & \textbf{104} & \textbf{161} & \textbf{4.44} \\
\bottomrule
\end{tabularx}
\end{table*}

\subsection{Session Types}
To better understand the value of traditional paper presentations and social events of conferences, we asked the participant to indicate the session type. \Cref{tab:surveyresults} depicts the complete numbers derived from our survey and the calculated average rating (arithmetic mean) for each session type. From the 13 pre-defined types as specified above, 7 types have been selected, while 5 participants indicated a \textit{other} type. Most attendees (17 + 44 \textit{ICSE main} = 61) evaluated \textbf{paper-presentations} and ranked these with 4.71 stars and 4.24 stars \textit{ICSE main} on average. The second most evaluated type are \textbf{co-located events} with 37 filled out surveys and an average ranking of 4.14 stars. The third most evaluated session type are \textbf{workshops} together with \textbf{keynotes}. With 23 responses, the workshops received an average rating of 4.87 stars, while keynotes are ranked with 5 stars and 4.13 stars \textit{ICSE main}.

\subsection{Other Feedback}
We also offered the option to further elaborate on the participant’s choice. Most feedback relates to the venue or organizational matters, as the following extracts show:

\begin{quoting}
"venue support can be much better." \\ \textbf{Session}: co-located event; \textbf{Stars}: 5
\end{quoting}

\begin{quoting}
"Rooms are too cold" \\ \textbf{Session}: workshop; \textbf{Stars}: 5
\end{quoting}

\begin{quoting}
"Why don’t you provide enough microphones?? It's hard to listen" \\ \textbf{Session}: paper-presentation session (ICSE main); \textbf{Stars}: 4
\end{quoting}

When we look deeper into the provided feedback, we see that participants tend to be more positive about the social sessions compared to the traditional paper presentations as the following extracts show:

\textbf{Positive attitude towards discussions and social interactions:}

\begin{quoting}
"Lots of conversation , fabulous discussions" \\ \textbf{Session}: Women@ICSE Lunch; \textbf{Stars}: 5
\end{quoting}

\begin{quoting}
"It is a nice Workshop" \\ \textbf{Session}: keynote; \textbf{Stars}: 5
\end{quoting}

\begin{quoting}
"Very engaging and hands-on `Work'shop :-)" \\ \textbf{Session}: paper-presentation; \textbf{Stars}: 5
\end{quoting}

\begin{quoting}
"Great GREENS!" \\ \textbf{Session}: paper-presentation; \textbf{Stars}: 5
\end{quoting}

\begin{quoting}
"This was student mentoring. I was a mentor  I really enjoyed it." \\ \textbf{Session}: workshop; \textbf{Stars}: 5 
\end{quoting}

\begin{quoting}
"I will like to see more of this workshop in the future. I learn a lot about STGT and will recommend this workshop to continue." \\ \textbf{Session}: workshop; \textbf{Stars}: 5
\end{quoting}

\textbf{Negative attitude towards traditional presentations:}

\begin{quoting}
"Quality of presentations was quite mixed, having very good, but also substandard ones." \\ \textbf{Session}: paper-presentation session (ICSE main); \textbf{Stars}: 3
\end{quoting}

\begin{quoting}
"I think generally it would be nicer to have more discussion than presentations" \\ \textbf{Session}: co-located event; \textbf{Stars}: 3   
\end{quoting}

\begin{quoting}
"Too many presenters with such poor English they can barely understand questions :/" \\ \textbf{Session}: paper-presentation session (ICSE main); \textbf{Stars}: 3
\end{quoting}

\begin{quoting}
"Unfortunately, some [removed] speakers are very hard to understand and i struggle to follow presentations. I do think their works are excellent, making it even more frustrating to not understand (not a native English speaker myself makes it even harder)." \\ \textbf{Session}: paper-presentation session (ICSE main); \textbf{Stars}: 1
\end{quoting}

\subsection{Carbon Footprint}
We estimate the overall footprint from and to Melbourne (return flight) of our 1,424 conference attendees to around \textbf{5,043.5 tonnes of CO2e}, which is equal to
the electricity required to power around 1,000 homes in a year \footnote{Greenhouse Gas Equivalencies Calculator - \url{https://www.epa.gov/energy/greenhouse-gas-equivalencies-calculator}}. The final calculation data in the form of a spreadsheet is available online\footnote{Carbon footprint data - \url{https://docs.google.com/spreadsheets/d/15yWv_niENp-FR79evvoHkau1QIiZLP9CnBbzBBtI17Y}}. As mentioned, this calculation can only be considered as an estimate due to several assumptions (e.g., nearest airport to given address), simplifications (e.g., one-leg flights), and inaccuracies (e.g., wrong airport code calculated by API).

\section{Conclusion}
By shedding light on the relationship between environmental considerations and perceived value in academic conferences, this report aimed to contribute to the ongoing dialogue surrounding sustainable practices in academic conferences. Unfortunately, the low response rate of our study limits possibilities to derive significant insights and conclusions. Furthermore, it is important to acknowledge the potential presence of question-bias in our survey. Respondents may have answered the survey questions based on their perception of the overall conference quality, rather than explicitly expressing their preferences regarding future session settings. Nevertheless, while the conference of the ICSE edition 2023 in Melbourne, Australia was rated with more than 4 stars on average, some open feedback comments lead us to the conclusion that traditional paper-presentations are less appreciated among the conference participants compared to the social events like workshops and networking sessions.

In order to get a better response rate in the future and prevent the potential question bias, we want to improve our survey twofold: (i) the survey should be as convenient as possible, hence, using a "smiley face survey kiosk" systems\footnote{Survey Kiosk system - e.g., \url{https://www.surveystance.com/smiley-face-survey-kiosk-app/}} might be a better choice compared to an online questionnaire; (ii) the survey goal needs to be better communicated and announced. Despite the low response rate and the rather less insights, we believe that such a survey can serve as a starting point for future surveys to evaluate the value of traditional academic conferences in the field of software engineering.

Our carbon footprint calculation estimates the CO2e associated with air travel to and from Melbourne, Australia to be approximately 5,043.5 tonnes of CO2e. It should be noted that this is an estimate only and does not include other factors such as ground travel or emissions related to the conference venue, catering or accommodation \cite{FunkeLago_LetStart_2022}. This estimate, however, can be used as a baseline for future conference editions in different countries and continents to establish historical data over time. To improve and simplify the calculation of air travel emissions, we suggest including a carbon footprint calculator in the conference registration form. Participants will be able to indicate their actual origin, preferred airport and stopover. In addition, such integration would raise awareness by presenting the footprint directly to the conference attendee during registration.

\bibliographystyle{plain}
\bibliography{biblio}

\end{document}